\documentclass[aps, preprint]{revtex4}  
\usepackage{graphicx}  
\usepackage{dcolumn}   
\usepackage{bm}        
\usepackage{amssymb}   

\begin{document}



\title{Comment on ``Finite Conductance Governs the Resonance Transmission of Thin Metal Slits at Microwave Frequencies"}
\author{Reuven Gordon \thanks{rgordon@uvic.ca}}
\author{Shuwen Chen}
\affiliation{Department of Electrical and Computer Engineering, University of Victoria, Victoria, British Columbia, V8P 5C2, Canada}


\maketitle
In the letter ``Finite Conductance Governs the Resonance Transmission of Thin Metal Slits at Microwave Frequencies" \cite{bb1}, the authors compare electromagnetic simulations and experiments to an analytic expression for the resonant transmission through a slit in a metal film provided by Eq. 9 of Ref. 2.  Eq. 9 does not represent accurately the locations of the resonances, and an improved fit can be found from the maxima of Eq. 8 of that same paper \cite{bb2}, as shown in Fig.~\ref{fg1} below. The analytic expression of the transmission found from single mode matching as Eq. 4 in Ref. 3 for normal incidence:
\begin{equation}
\label{eq1}
T = \frac{{{{\left| t \right|}^2}\left( {1 - {{\left| r \right|}^2}} \right)}}{{{{\left| {1 - {r^2}\exp \left( {i4\pi l} \right)} \right|}^2}}}
\end{equation}
with $t = {2 \mathord{\left/
 {\vphantom {2 {\left( {1 + t} \right)}}} \right.
 \kern-\nulldelimiterspace} {\left( {1 + t} \right)}}$, $r = {{\left( {1 - I} \right)} \mathord{\left/
 {\vphantom {{\left( {1 - I} \right)} {\left( {1 + I} \right)}}} \right.
 \kern-\nulldelimiterspace} {\left( {1 + I} \right)}}$ and $I = a\pi  + 2ai\left[ {\ln \left( {2\pi a} \right) - 3/2} \right]$, and where $a$  and $l$  are the slit width and thickness, normalized to the wavelength of the incident light. This gives improved agreement, as shown in Fig.~\ref{fg1}.

The formulation of Eq.~\ref{eq1} above can account for the finite conductivity. If we consider that the propagation constant is corrected in the slit by the finite conductivity using the usual expression for the lowest order TM mode \cite{bb4}, an effective increase in the length of the slit can be found as:
\begin{equation}
\label{eq2}
\Delta l = \frac{l}{{\sqrt 2 \pi {{\left| {{\varepsilon _{\text{m}}}} \right|}^{1/2}}a}} 
\end{equation}

where ${\varepsilon _{\text{m}}}$ is the relative permittivity of the metal (mainly imaginary for the microwave region). Using this correction the single mode matching theory can fit the experimental data, as shown in Fig.~\ref{fg2} below.

In summary, the results provided by matching the lowest order slit mode to the continuum of free space modes provides quantitative agreement with the experiment of Ref. 1, and can account for the effects of the finite conductivity with a small modification to the effective wavelength.

 \begin{figure}
\includegraphics[scale=0.45]{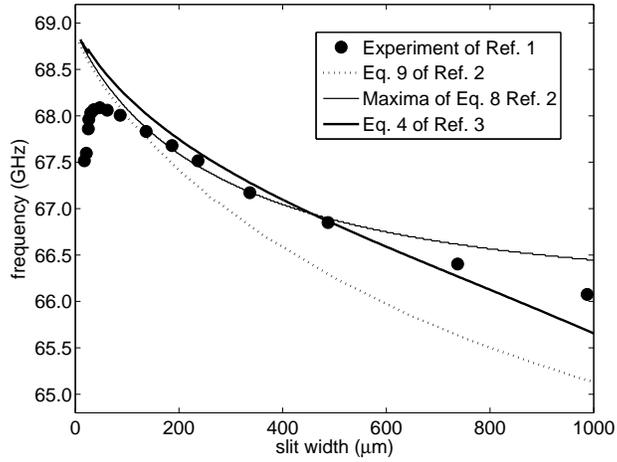}
\caption{\label{fg1} Transmission resonance frequency found in experiment of Ref. 1, as compared with analytic expression of Eq. 9 in Ref. 2, the maxima from Eq. 8 in Ref. 2, and the analytic expression of Eq. 4 in Ref. 3.}
\end{figure}

\begin{figure}
\includegraphics[scale=0.45]{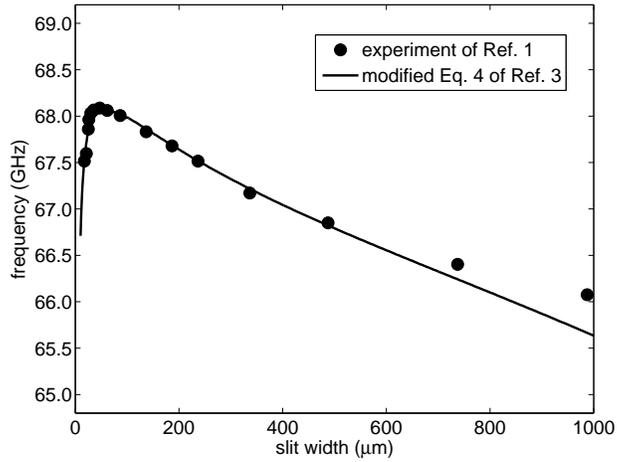}
\caption{\label{fg2} Transmission resonance frequency found in experiment of Ref. 1, as compared with analytic expression of Ref. 3, modified to include the effect of finite conductivity by scaling the length of the slit.}
\end{figure}

\end{document}